\documentclass[aps, twocolumn, showpacs, letterpaper]{revtex4}

\usepackage{color}
\usepackage{amsmath}
\usepackage{amssymb}
\usepackage{graphicx}
\usepackage{xspace}

\renewcommand{\exp}[1]{\textrm{exp}\left(#1\right)}

\begin{document}

\title{Suppression of Tearing Modes by RF Current Condensation}

\author{A.~H.~Reiman and N.~J. Fisch}
\affiliation{Princeton Plasma Physics Laboratory, Princeton University, Princeton, New Jersey 08544, USA}
\date{\today}

\begin{abstract}
Currents driven by  rf (radio frequency) waves in the interior of magnetic islands can stabilize deleterious tearing modes in tokamaks. 
Present analyses of stabilization  assume that the local electron acceleration
is unaffected by the presence of the island. 
However, the power deposition and electron acceleration are sensitive to the perturbation of the temperature.  The nonlinear feedback on the power deposition in the island increases the temperature perturbation, and can lead to a bifurcation of the solution to the steady-state heat diffusion equation.  The combination of the nonlinearly enhanced temperature perturbation with the rf current drive sensitivity to the temperature leads to an rf current condensation effect, which can increase the efficiency of rf current drive stabilization and reduce its sensitivity to radial misalignment of the ray trajectories. The threshold for the effect is in a regime that has been encountered in experiments, and will likely be encountered in ITER.

\end{abstract}

\maketitle
\date{\today}

\maketitle

{\it Introduction:} \
A study of the root causes of disruptions in the JET tokamak found that neoclassical tearing modes (NTMs) were the single most common cause \cite{devries,devries14}. Theoretical calculations in the early 1980's showed the feasibility of using rf current drive to stabilize tearing modes \cite{reiman,Yoshioka}. The recognition in the late 1990's that bootstrap currents were driving NTMs in hot, collisionless tokamak plasmas \cite{Chang,LaHaye97,ZohmEPS97,Gates97}, led to a resurgence of theoretical work in this area \cite{hegna,ZohmPOP97,Harvey_Perkins,Yu2000,Prater03}, to experimental demonstrations of stabilization \cite{Bernabei,Warrick2000,Gantenbein,Zohm2001,Isayama2000,LaHaye2002,Petty04},
and to continuing intensive attention  
\cite{Sauter2004,Kamendje2005,LaHaye2006,Lahaye08,Henderson08,Lazzari,Volpe,Sauter2010,Bertelli2011,Hennen2012,Smolyakov_2013,Ayten,Borgogno,Volpe2015,Fevrier2016,Wang2015,JC_Li_2017,
Grasso_JPP2016,Grasso18,Poli15}. 
A variety of rf waves are used to drive current \cite{fisch87}, but, for stabilizing the NTM, the most studied methods are electron cyclotron current drive (ECCD)  \cite{fisch80} and  lower hybrid current drive (LHCD) \cite{fisch78}.  
ITER is designed with an NTM ECCD stabilization 
capability,
with continued effort to model and  improve  this capability \cite{Henderson08,Hennen2012,Bertelli2011,Poli}.
We identify here an rf current condensation effect, previously overlooked, which can significantly facilitate island stabilization.

Calculations of rf stabilization of magnetic islands assume, at present, that the local acceleration of electrons is unaffected by the presence of the island. 
However, the local deposition is sensitive to 
small changes in the temperature, and 
these changes 
can be significantly affected by the presence of an island. The effect on the local deposition becomes significant when the fractional temperature perturbation exceeds about 5\% for electron cyclotron waves and 2.5\% for lower hybrid waves. Temperature perturbations as high as 20\% have been measured in islands in rf stabilization experiments \cite{Westerhof07}.

In the conventional picture of rf current drive stabilization of a rotating island, a geometric effect associated with the equilibration of the rf driven current density within the flux surfaces of the island leads to a higher current density near the center of the island than near its periphery, and to  a corresponding stabilizing resonant component of the field. We show that the sensitivity of the current drive and power deposition to small changes in the temperature can give rise to a ``current condensation'' effect that can greatly 
concentrate
the current density near the center of the island, 
thereby greatly increasing the efficiency of the stabilization. 
Thus,  a given rf power stabilizes larger islands.

Conventional stabilization by rf driven currents is sensitive to the radial alignment of the current deposition profile with the O-line (center) of the magnetic island. The effect changes sign (becoming destabilizing) if the current deposition is displaced a distance $0.5 \max(W_i, W_d)$ relative to the O-line, where $W_i$ is the island width and $W_d$ is the width of the deposition profile \cite{Lazzari}. 
Current condensation reduces the sensitivity of the stabilization to precise alignment of the RF ray trajectories.
Even a broad rf driven current, primarily for steady state  operation, condenses, thereby providing stabilization even  absent a stabilizing geometric effect.

{\it Power  and Current Deposition:} \ 
The power deposition by electron cyclotron \cite{karney81} and lower hybrid waves \cite{karney79} is sensitive to the temperature because they deposit their energy on the electron tail.  Let $v_0$ be the the electron speed at the location in velocity space of greatest power deposition. The deposition in that region is proportional to the number of electrons there, $P_{rf} \propto \exp{-w^2}$, where $w\equiv v_0/v_T$,    $mv_T^2/2 = T$, and $T$ is the electron temperature. For a small temperature perturbation, $\tilde T$, the change in the local power deposition produced by the perturbation is given by
\begin{equation} \label{Prf}
P_{rf} \propto \exp{-w^2} = \exp{ -w_0^2} \exp{w_0^2  \tilde T/T_0},
\end{equation}
where $T_0$ is the unperturbed temperature and $w_0$
is the value of $w$ in the absence of the temperature perturbation. 
Typically $w_0^2 \approx 10$ for ECCD and $w_0^2 \approx 20$ for LHCD. The power deposition is thus sensitive even to a small $\tilde T / T_0$, even as other quantities, such as the dispersion relation, are not. The rf driven current similarly grows exponentially, but with $w_{rf}^2 \tilde T / T_0$, where $w_{rf}$ is the resonant velocity producing the maximum current.  For high current drive efficiency, $w_0 \approx w_{rf}$, with $w_0$ substantially in the parallel direction.  For simplicity, in the following we will assume $w_0=w_{rf}$.

We consider in turn the two pieces to the current condensation effect: the increase of the rf current with increasing temperature, and the nonlinear feedback arising from the increased power deposition with increasing temperature, which enhances the temperature perturbation.

{\it Sensitivity of Current Density to Temperature:} \ 
Both the ohmic current and the rf driven current are affected by the temperature, which is peaked at the O-line because of the well known effect of the thermal insulation in the island. 
The effect of the Spitzer ohmic current perturbation has been extensively studied, and it is believed to have provided a significant stabilizing effect in a number of experiments
%
  \cite{Kurita94,hegna,Westerhof07,Lazzari,Maget18,MagetHAL}.
  There is experimental evidence of strongly reduced transport in the interior of islands \cite{Inagaki,Ida,Bardoczi,Bardoczi17}, and the associated increase in the temperature perturbation will enhance both effects.

The Spitzer current density perturbation $\Delta J$ produced by a temperature perturbation $\tilde T$ is
$
\Delta J_{\rm Sp} / J_{\rm Sp} =  \Delta\sigma_{Sp} / \sigma_{Sp} = (3 / 2) \tilde T / T_0.
$
It follows from Eq. (\ref{Prf}) that the perturbation of the rf driven current is
$
\Delta J_{rf} / J_{rf} \approx \exp{w_0^2 \tilde T/T_0} - 1 > w_0^2 \tilde T/T_0.
$
The perturbation of the rf driven current 
 can dominate that of the ohmic current density even when the unperturbed rf driven current density is relatively small. 
When the bootstrap current density is comparable to the ohmic current density, as it is expected to be at the $q=2$ surface in ITER, the rf current density needed for NTM stabilization is comparable to the ohmic current density.

Some implications of the rf current density dependence on temperature for the rf current drive stabilization of magnetic islands are discussed in 
Ref.~\cite{reiman}.  Although the discussion of the effect there is in the context of LHCD, it is only assumed that the power is deposited on the electron tail, so the calculations there apply also to ECCD.  Although the ohmic effect continues to be the subject of intensive research \cite{Maget18,MagetHAL}, the effect of the rf current perturbation has not been investigated beyond the calculations of Ref. \cite{reiman}. Here we show that the combination with the nonlinear self-reinforcement of the temperature perturbation leads to the rf current condensation effect.

{\it Nonlinear Feedback Effect on the Temperature:} \ 
The considerations of this section will be applicable to ECH or lower hybrid heating in an island, regardless of whether there is unidirectional injection for current drive, and will therefore be applicable also to ohmic stabilization.

For an NTM, the temperature in the island equilibrates on a time short compared to the growth time of the island, suggesting that we consider
the steady-state diffusion equation
$
\nabla  \cdot \left ( n {\bf \kappa} \cdot \nabla \tilde T \right ) = - P_{rf},
$
where $n$ is the density and ${\bf \kappa}$ is the thermal conductivity tensor. 
We assume that the island is sufficiently large that the temperature is constant within the flux surfaces in the island  \cite{Fitzpatrick,Hazeltine-transport_1997}.  The unperturbed temperature ($P_{rf}=0$) is flat in the island. For simplicity, we
take $n$ and the perpendicular thermal diffusivity, $\kappa_{\perp}$, to be constant in the island.

Consider the case where $\tilde T/T_0$ is small, but $w_0^2 \tilde T/T_0$ is not necessarily small. We are interested in the temperature in the island interior relative to that at the separatrix, and we can set $T_0 = T_s$, where $T_s$ is the temperature at the separatrix, absorbing a constant factor $\exp{w_0^2(T_s-T_0)/T_0}$ into $\bar{P}_0$ below. For typical ECCD applications, the change of the wavenumber in the island is small, giving $w_0 \approx w_s$, where $w_s$ is the value of $w_0$ at the separatrix. Neglecting wave depletion, and using 
Eq.~(\ref{Prf}), we can write $P_{rf} = \bar{P}_0(\rho) \exp{w_s^2 \tilde T/T_s}$.  We take $\rho = 0$ at the O-line.   
We consider the case where $\bar{P}_0$ is independent of $\rho$, corresponding to an unperturbed power deposition profile broad compared to the width of the island.  (The power deposition outside the island does not affect the temperature perturbation inside.)


Consider first 
a simple slab model, which can be solved analytically, with $x=0$ representing the O-line and $x=\pm W_i / 2$ representing the separatrix.  Letting $u \equiv w_s^2 \tilde T/T_s$ and $P_0 \equiv W_i^2 w_s^2 \bar{P}_0/ \left( 4 n \kappa_{\perp} T_s \right)$, where $W_i$ is the island width, and normalizing $x$ to the island half-width, the diffusion equation becomes $d^2 u / d x^2= -P_0 \exp{u}.$
We solve this equation explicitly, getting
$
u(x)=\ln \left( \lambda_1 / 2 P_0 \right)
	-2 \ln \left\{ \cosh \left[ \sqrt{\lambda_1} \left( x - \lambda_2 \right) / 2 \right] \right\},
$
where $\lambda_1$ and $\lambda_2$ are constants of integration.  The boundary conditions, $d u / d x =0$ at $x=0$ and $u=0$ at $x=1$, yield a nonlinear eigenvalue equation
$
\lambda_1 = 2 P_0 \cosh^2 \left( \sqrt{\lambda_1} / 2 \right).
$
It has two roots below a threshold in $P_0$ corresponding approximately to $P_0 = 0.88$, and no roots above that threshold.  This is a fold bifurcation.  In the context of catastrophe theory, this type of behavior is known as a {\it fold catastrophe}  \cite{Arnold}.

\begin{figure}[h!]
\includegraphics[width=0.5\textwidth]{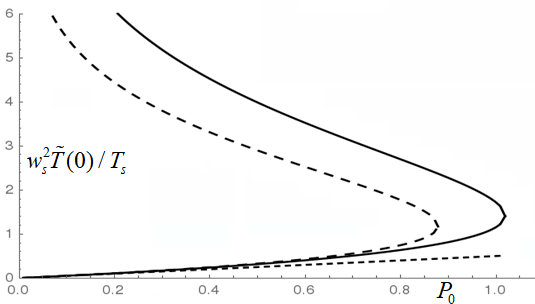}
\caption{ Temperature differential $w_s^2 {\tilde T}(0) / T_s$ vs.~normalized  power density coefficient $P_0$, showing the fold bifurcation. }
\end{figure}

\vspace{-0.5em}
The value of $w_s^2 \tilde T / T_s$ at $x=0$ as a function of $P_0$ is shown as the inner dashed line in Fig.~1. 
Above the threshold value of $P_0$, there is no steady-state solution for small $\tilde T(0) / T_s$.
The temperature in the island increases until it becomes large enough to encounter additional physics, such as the depletion of the energy in the rf wave, or a gradient threshold for stiff transport.

Calculated solutions to the diffusion equation for a range of $P_{rf}$ profiles in the slab, in addition to a constant $P_{rf}$, find that that bifurcation threshold is relatively insensitive to the width of the profile. 

The solid line in Fig.~1 corresponds to the solution for a more accurate treatment of the diffusion, taking into account the geometry of the island flux surfaces, with a uniform unperturbed power deposition in the island.  
%
This treatment employs a conventional cylindrical model for the magnetic field,
$
{\bf B} = \nabla \psi \times \hat{z} - (k r / m)B_z \hat{\theta} + B_z \hat{z},
$
where we can expand $\psi$ about the rational surface as
$\psi = \psi_0'' x^2/2 - \epsilon \cos(m \zeta)$,
$\zeta =  \theta - k z / m$, and $\epsilon$ is a constant (the ``constant-psi approximation'') \cite{White,Fitzpatrick}.
We define $\rho^2 = \psi / 2 \epsilon +1/2$.
After some algebra, and after discarding a term small in $W_i/R$, where $R$ is the major radius, the diffusion equation takes the form
\vspace{-0.5em}
\begin{equation*}
\frac{d}{d \rho } \left( \frac{E(\rho ) - \left( 1 - \rho ^2 \right)K(\rho )} {\rho } \frac{d}{d \rho }u\left( \rho  \right)  \right)
= P_0 \rho K(\rho) \exp{u},
\end{equation*}
where $K(k) \equiv \int_0^{\pi/2}  {{{(1 - {k^2}{{\sin }^2}\chi)}^{ - 1/2}}d\chi}$ is the complete elliptic integral of the first kind, and $E(k) \equiv \int_0^{\pi/2}  {{{(1 - {k^2}{{\sin }^2}\chi)}^{1/2}}d\chi}$ is the complete elliptic integral of the second kind. The boundary conditions are $u=0$ at the separatrix and $du / d \rho = 0$ at the O-point.
The bifurcation threshold corresponds to $P_0 \approx 1.02$. 

The bottom curve in Fig.~2 shows the solution of the linear diffusion equation, which neglects the dependence of the power deposition on the temperature.

The bifurcated solution corresponds to the following physical picture. 
Initially, at low $\tilde T$, the power deposition term in the time-dependent heat diffusion equation dominates, and the temperature increases.  
The second derivative increases with increasing temperature, until it balances the power deposition at the lower root of the steady-state diffusion equation.  
A perturbation to a higher temperature gives a further increase in the second derivative, so that the lower root is stable.  
At sufficiently high temperature, the exponential begins to dominate, and the power deposition increases more rapidly with increasing temperature until the two terms again balance at the second root.  
The power deposition continues to increase more rapidly with increasing temperature, so that the second root is unstable. 
The temperature then continues to increase until 
limited by an effect not considered here, such as those mentioned above, giving a third, stable solution branch. 
The two lower solution branches merge at the bifurcation point.  
Above the bifurcation point, the increase of the power deposition with temperature begins to dominate before a balance with the diffusive term is reached, and the temperature rises until the uppermost solution branch is reached. Interestingly, if the island width is now decreased, there is a hysteresis effect, with the solution moving along the uppermost branch, leading to smaller saturated island widths.

{\it Increased Stabilization Efficiency and Decreased Sensitivity to Alignment of Ray Trajectories:} \
The exponential dependence of the driven current on the temperature combines with the nonlinear effect on the temperature perturbation to give an rf current condensation effect.
A widely used measure of the efficiency of RF current drive stabilization is the ratio of the resonant Fourier component of the current to the total RF driven current:
$
\eta_{\rm stab} = \int_{-\infty}^{\infty}dx\oint d\zeta  j_d cos(m \zeta) \Big/ \int_{-\infty}^{\infty}dx\oint d\zeta  j_d
$
\cite{hegna,Giruzzi99,Zohm07,Poli15}. (The quantity $\Delta'$ in the {\it modified Rutherford equation} \cite{White,reiman,hegna,LaHaye2006} is proportional to the resonant component of the current.) Using the temperature profiles calculated in the previous section, we calculate the efficiency for a broad, Gaussian deposition profile, $P_{rf} =  \bar{P}_0 e^{-4x^2/W_d^2} \exp{w_s^2 \tilde T/T_s}$, with $W_d \gg W_i$. We again define   $P_0 \equiv W_i^2 w_s^2 \bar{P}_0/ \left( 4 n \kappa_{\perp} T_s \right)$.  We find
$
\eta_{\rm stab} = \eta_0 [1 + (W_d/W_i) R(P_0)],
$
where $\eta_0=0.25 (W_i/W_d)^2$ is the conventionally calculated efficiency \cite{Lazzari}, associated with the geometric effect, and $R(P_0)$ is shown in Fig.~2.

The current condensation contribution to the efficiency dominates when $(W_d/W_i) R(P_0) > 1$. Approaching the bifurcation threshold, the current condensation contribution to the efficiency is $O(W_d/W_i)$ times $\eta_0$. 
The stabilizing effect is relatively insensitive to the radial alignment of the ray trajectories as long as the contribution of the current condensation to $\eta_{\rm stab}$ dominates the contribution from the geometric effect.
\begin{figure}[h!]
\includegraphics[width=0.5\textwidth]{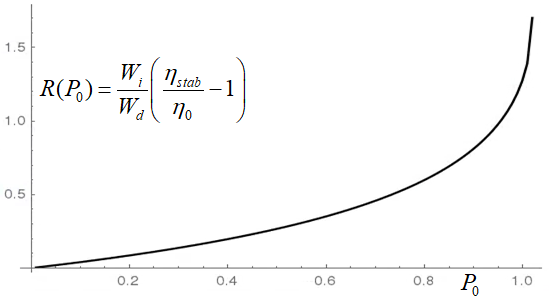}
\caption{$(W_d/W_i) R(P_0)$ is the relative magnitude of the contribution of the condensation effect to the stabilization efficiency for broad, Gaussian deposition profiles, where $W_d$ is the width of the deposition profile and $W_i$ is the island width.}
\end{figure}

\vspace{-1em}
{\it Experimental Relevance:}
When $w_s^2 \tilde T/T_s = 0.5$, there is a 65\% increase in the local power deposition, and a larger increase in the rf current density, relative to the conventionally calculated values.  
Above this level of temperature perturbation, there is an exponential increase.
Tearing stabilization experiments via electron cyclotron waves on TEXTOR  found $\tilde T(0)/T_s \approx 0.2$ \cite{Westerhof07}.
ASTRA transport simulations for an ITER 2/1 magnetic island in Ref. \cite{Westerhof07} considered a 24 cm island with 20 MW of heating power, 
finding $\tilde T(0)/T_s \approx 25\%$ for $\chi_e = 0.1 m^2/sec$. The linear $\tilde T(0)/T_s$ scales as $W_i P_{tot}$, where $P_{tot}$ is the total power deposited in the island, implying that the local enhancement will become significant when the island width is about 5 cm for 20 MW of heating power, or 10 cm with 10 MW.

On ITER, it is important to minimize the ECCD power usage for NTM stabilization.
That, together with low predicted thresholds for island locking, has led to scenario studies that envision the stabilization of islands at small widths, with modest power deposition \cite{Zohm07,Henderson08,Hennen2012,Bertelli2011,Poli}. 
There will be a tradeoff between the desire for small deposition widths to minimize the required power, and the risk of misalignment. 
When the island widths remain small and the ECCD power deposition in the islands is modest, the rf current condensation effect will not come into play.  
It can be anticipated, however, that there will be off-normal events, 
such as flakes falling into the plasma or abnormally strong sawtooth events, etc., so that the attempted stabilization at small island widths will not be 100\% successful.  
It will be critical to stabilize the resulting large islands to prevent disruptions, using whatever power is available. 
The current condensation effect can 
then be crucial.

The experimental bifurcation threshold can be estimated from
our calculation of the nonlinear enhancement of $\tilde T$ in an island, yielding $\tilde T(0)/T_s \approx 0.14$ for ECCD and $\tilde T(0)/T_s \approx 0.07$ for LHCD. The observed  $\tilde T(0)/T_s \approx 0.2$ on TEXTOR suggests that the experiment may have approached or exceeded the bifurcation threshold.  
Exceeding the bifurcation threshold 
gives a hysteresis effect, with the island suppressed to widths below what would otherwise be achievable.  Suggestively,  the  experiment observed suppression to widths well below the calculated widths of the deposition profiles, where the geometric stabilizing effect is predicted to be much reduced \cite{Westerhof}.

{\it Hot Conductivity Current Condensation and Destabilization:} \  
In the presence of rf current drive, the ohmic current can be written as $J_{\rm OH} = J_{\rm Sp} + J_{\rm H}$,  where $J_{\rm Sp} = \sigma_{Sp}E$ is the ohmic (Spitzer) current in the absence of the rf, 
and $J_{\rm H} = \sigma_{H}E$ is due to the {\it hot conductivity} $\sigma_{H}$, arising from electron velocity space distortions proportional to the rf power dissipated \cite{fisch85a}. Although the hot conductivity current is relatively small for usual tokamak operation, it may play a critical role in the case of rf current overdrive, which occurs when the rf is utilized for start-up operation \cite{fisch85b,leuterer,Giruzzi97,takase,chen,ding},
or when it is oscillated  to optimize the current drive efficiency \cite{fisch_transformer,Li_2012}. The hot conductivity  $\sigma_{H}$ has been theoretically predicted \cite{fisch85a} and experimentally verified in detail \cite{karney_fisch_jobes}.  It is proportional to $P_{rf}$, so it displays the same exponential sensitivity to temperature perturbations as the rf driven current, and the same current condensation effect.


However, in exceeding  the total toroidal current during rf current overdrive, the  rf-driven current induces  a toroidal electric field that opposes the rf-driven current, with $J_{\rm OH} \simeq - J_{\rm rf}$.  
Now a change $\tilde T$ at the O-line produces a $\Delta J_{\rm OH}$ opposite to both the total current and the rf-driven current, and so is destabilizing rather than stabilizing. For strong overdrive, the ohmic countercurrent is mainly limited by the hot conductivity, with $J_{\rm OH} \simeq  J_{\rm H}$  \cite{fisch85a}.  
In contrast to the Spitzer current increment, $\Delta  J_{\rm H} $  nearly matches the rf incremental current $\Delta J_{rf}$, except that it is destabilizing.  
Moreover, both for LHCD and ECCD, it is inevitable that some rf  power will drive current opposite to $J_{\rm rf}$,  which will  further increase  $J_{\rm H}$ relative to $J_{\rm rf}$.  
%
Thus,  it will be more difficult to stabilize the NTM in the rf overdrive mode in the limit where the overdrive is strong. 
A weaker rf overdrive would reduce this destabilization.

{\it Comparison of  Current Drive Methods:} \ 
Although other means of noninductive current drive have been contemplated  for NTM stabilization,  the current condensation effect described here is only available
 for ECCD and LHCD, because  their damping decrements are highly sensitive to the electron temperature.  
This sensitivity is not available for current drive methods based on sub-thermal electrons \cite{fisch81a}, such as through Alfven waves,  or neutral beam current drive  \cite{ohkawa}, even if it could be relatively localized through minority species heating \cite{fisch81b}.  

Most of the experimentation to date has involved ECCD rather than LHCD, perhaps in part because of the thought that ECCD could be better localized.
However, with current condensation, this localization may no longer be critical.
Also, launching  lower hybrid waves from the tokamak high-field side allows greater localization through single-pass absorption, and enables  
high-magnetic field compact tokamaks \cite{Wallace,Sorbom}.
In addition, LHCD, but not ECCD, can tap the  energy in  $\alpha$ particles  in a reactor  through the $\alpha$-channeling effect \cite{fisch1992interaction}, reducing the recirculating power. 
The channeling is in fact most effective under high-field side launch   \cite{ochs_2015a,ochs_2015b}.
LHCD also sees a stronger rf current condensation effect than ECCD, because of its higher phase velocity, a potential advantage for stabilizing NTMs.

{\it Discussion and Conclusions:} \
%
%
%
%
The rf current condensation effect identified here increases the efficiency of stabilization, so that larger islands can be stabilized for a given rf power.
It also reduces the sensitivity of the stabilization to radial misalignment of the ray trajectories relative to the island O-line.  
Even very broad rf-driven  currents, such as used for  steady state operation, can condense in large islands, thereby stabilizing them.

The local power deposition and electron acceleration are highly sensitive to the perturbation of the local temperature in an island.  
Moreover, the nonlinear feedback on the power deposition increases the temperature perturbation. The combination of the nonlinearly enhanced temperature perturbation with the rf current sensitivity to the temperature produces the rf current condensation effect.  

Our calculations here neglected the effects of wave depletion, which have been left for future investigation.  If account is taken of wave depletion in launching the rf waves, the effect can further increase the concentration of the rf current near the O-line, and can thereby further increase the stabilization efficiency. Also neglected were more peaked unperturbed deposition profiles and additional sources of heating in the islands, which would lower some of the thresholds calculated here.
  
Despite approximations, what is clear is that the current condensation effect is both new and important. 
Signatures of the phenomena predicted here  should be observable in more precise temperature measurements in island interiors, through comparisons of different methods of rf current drive, through more careful analyses of saturated island widths, and through comparisons to island formation in the rf overdrive regime. 
Apart from the academic interest of the fold bifurcation, it leads to the practical applications of increased stabilization efficiency, and decreased saturated island widths through hysteresis.
The threshold for the current condensation effect has been encountered in present-day experiments, and will very likely be encountered in ITER.
The condensation effect is particularly effective in stabilizing large islands, where the increased efficiency may be crucial for the minimization of disruptivity on ITER, which in turn could impact the economical advancement of tokamak fusion. 

{\it Acknowledgments:} \
The authors would like to acknowledge conversations with Ms.~Ge Dong and Mr.~Eduardo Rodriguez.  This work was supported by  DOE  Contract No.~DE-AC02-09CH11466. 


\end{document}